\documentclass[11pt,Arial]{article}

\usepackage[margin=1.0in]{geometry}
\usepackage{fancyhdr}
\usepackage{epsf}
\usepackage{graphicx}
\usepackage{amssymb,amsmath}
\usepackage[numbers,sort&compress]{natbib}
\usepackage[colorlinks=true,citecolor=blue,urlcolor=blue]{hyperref}
\usepackage{wrapfig}

\usepackage[bottom]{footmisc}
\usepackage[utf8]{inputenc}

\begin{document}

\begin{titlepage}
\includegraphics[width=150mm]{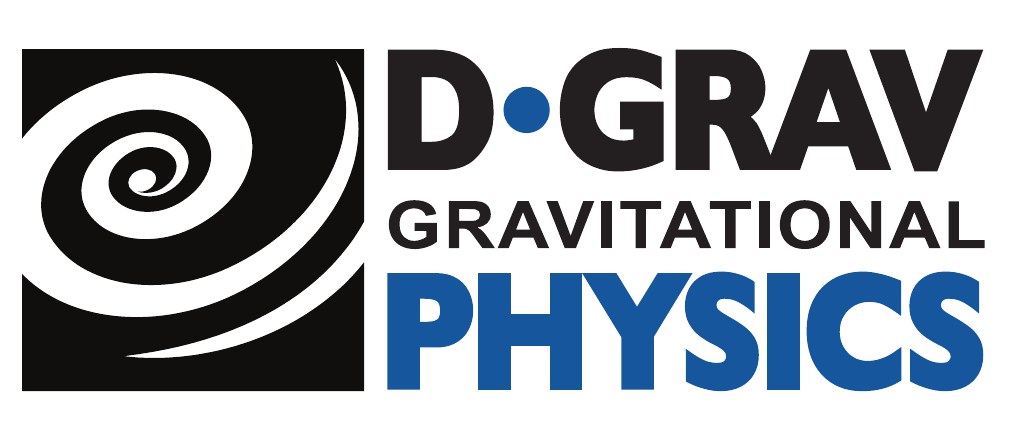}
\begin{center}
{ \Large {\bf MATTERS OF GRAVITY}}\\ 
\bigskip
\hrule
\medskip
{The newsletter of the Division of Gravitational Physics of the American Physical 
Society}\\
\medskip
{\bf Number 55 \hfill April 2022}
\end{center}
\begin{flushleft}
\tableofcontents
\end{flushleft}
\end{titlepage}
\vfill\eject
\begin{flushleft}
\section*{\noindent  Editor\hfill}
Aaron Zimmerman\\
\smallskip
Center for Gravitational Physics,
University of Texas at Austin\\
aaron.zimmerman@utexas.edu

\section*{Contributors to this issue}
Varun Srivastava (Syracuse University)

Kathryne J. Daniel (Bryn Mawr College)

Stefan Ballmer (Syracuse University)

\section*{\noindent Division of Gravitational Physics (DGRAV) Authorities:}
Chair: Gabriela Gonzalez (04/21–04/22)
Louisiana State University

Chair-Elect: Jennie Traschen (04/21–04/22)
University of Massachusetts Amherst

Vice Chair: Thomas W Baumgarte (04/21–04/22)
Bowdoin College

Past Chair: Nicolas Yunes (04/21–04/22)
University of Illinois at Urbana-Champaign

Secretary/Treasurer: Jocelyn S Read (04/20–04/23)
California State University, Fullerton

Councilor: Manuela Campanelli (01/21–12/24)
Rochester Institute of Technology

Member-at-Large: Alessandra Corsi (04/19–04/22)
Texas Tech University

Member-at-Large: Henriette D Elvang (04/19–04/22)
University of Michigan

Member-at-Large: Sarah J Vigeland (04/20–04/23)
University of Wisconsin - Milwaukee

Member-at-Large: Matthew J Evans (04/20–04/23)
Massachusetts Institute of Technology MIT

Member-at-Large: Marcelle Soares-Santos (04/21–04/24)
University of Michigan

Member-at-Large: Kent Yagi (04/21–04/24)
University of Virginia

Student Member: Maya Fishbach (04/20–04/22)
Northwestern University

Student Member: Delilah E A Gates (04/21–04/23)
Princeton University

\bigskip

DISCLAIMER: The opinions expressed in the articles of this newsletter represent
the views of the authors and are not necessarily the views of APS.
The articles in this newsletter are not peer reviewed.
\end{flushleft}

\bigskip
\hfill ISSN: 1527-3431

\pagebreak

\section*{Editorial}
\addtocontents{toc}{\protect\medskip}
\addcontentsline{toc}{section}{\bf Editorial}

It has been an extraordinary couple of years, to say the least. 
During the global pandemic, each of us and our entire field faced unprecedented challenges.
Conferences went hybrid or fully virtual, Zoom and Slack became core tools for continuing collaborations, and the demands of caring for our families, friends, students, and ourselves were overwhelming for many.
I know I found it impossible to keep up with all my obligations, and one I sadly let drop was my promise to carry on as the editor of Matters of Gravity after being handed the reins by longtime editor David Garfinkle.
I failed miserably in that duty, for which I can only sincerely apologize to all of those who looked forward to this twice-yearly newsletter from the Division of Gravitational Physics (DGRAV) of the American Physical Society.
Apologize, and try to do better in the future.

Even now, after more than two years of pandemic, the future remains uncertain.
I am at least hopeful that we are moving towards a more normal time, and I'm looking forward both to the hybrid APS April meeting and the possibility of seeing old friends and new faces at summer conferences.
\begin{wrapfigure}{r}{0.25\textwidth}
\begin{center}
\vspace{-3mm}
\includegraphics[width=0.25\textwidth]{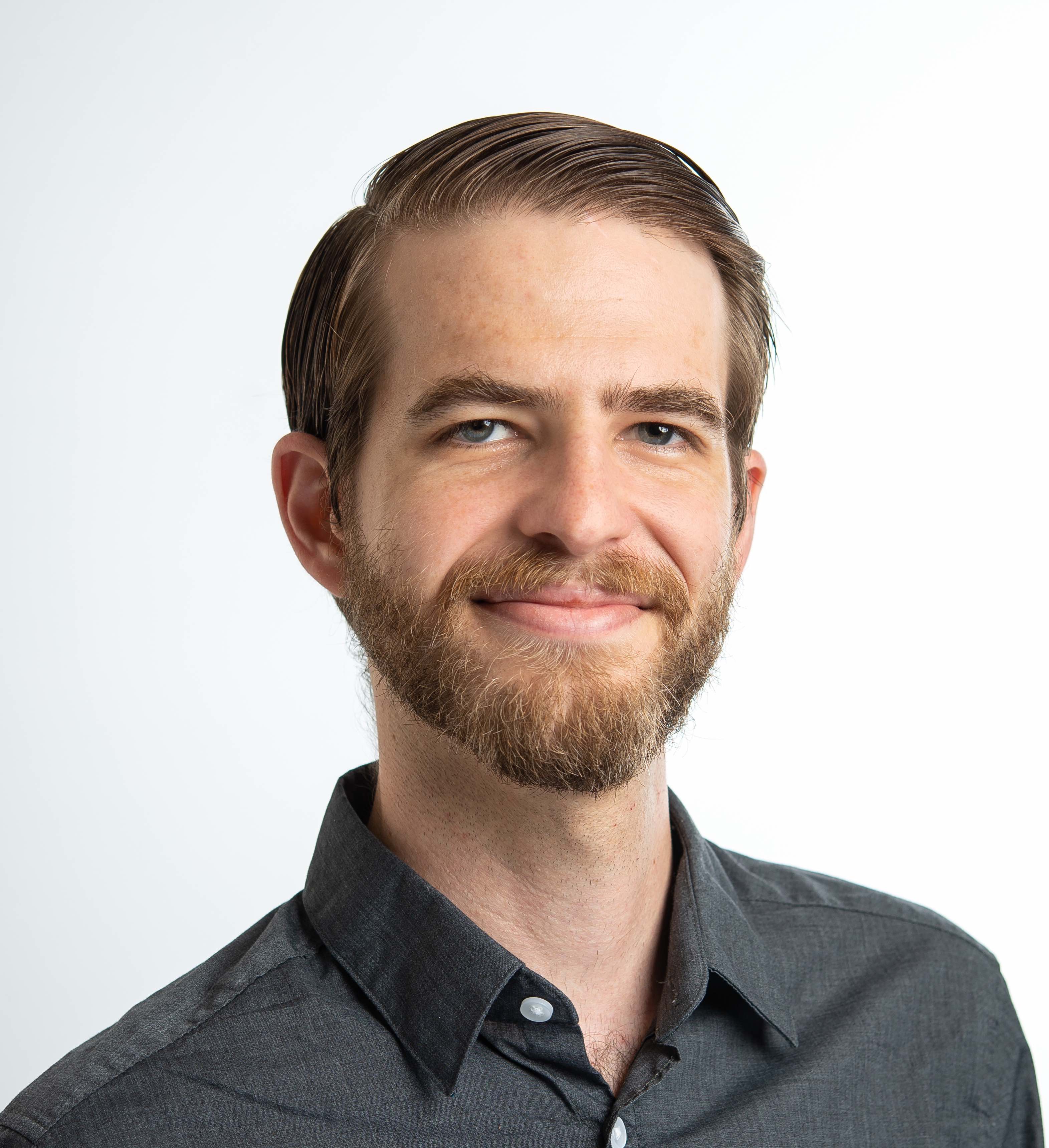}
\end{center}
\vspace{-3mm}
\end{wrapfigure}
For those that don't know me, let me take this opportunity to introduce myself.
I am currently an Assistant Professor at the University of Texas at Austin.
My PhD training at Caltech was mostly focused on black hole ringdown, supervised by Yanbei Chen. 
Since then I have also worked in numerical relativity and data analysis with the LIGO Scientific Collaboration.
I hope I can bring the perspective of both a theorist and an analyst of experimental data when soliciting articles and conference reports for Matters of Gravity.

I want to sincerely thank David Garfinkle for his years of hard work keeping Matters of Gravity going.
In this brief issue, meant to restart the circulation of this newsletter, you will find a schedule for DGRAV-sponsored activities at the coming April APS meeting in New York, and an article by Varun Srivastava, Kathryne J. Daniel, and Stefan Ballmer detailing the ambitious Cosmic Explorer Observatory concept, including a few words on how you can get involved.
The next newsletter will be targeted for December 2022, and then we will return to the twice-yearly cadence of June and December issues, including the usual content of news from DGRAV (We Hear That) and information about DGRAV activities and membership.
The December issues generally include information about the DGRAV-sponsored sessions for the next year's APS meeting (as you can see, this information is being provided by me quite late this year).
But first, an appeal to you, the reader:\\

\section*{Call for contributions}

Have you been to an interesting conference lately? 
Do you have some news to share with the community? 
Maybe you've noticed an interesting development in gravitation or a neighboring field, and would like to write a short summary article on the development?
If so, please reach out!
{\bf Matters of Gravity is seeking correspondents, new and old} for our next issues.
We are especially hoping to recruit correspondents to report on conferences and workshops, so please let us know if you are interested in contributing a brief meeting report to an upcoming issue. \\

\hfill -- Aaron Zimmerman

\pagebreak

\section*{DGRAV activities at the APS April Meeting}
\addtocontents{toc}{\protect\medskip}
\addcontentsline{toc}{section}{\bf DGRAV sessions at APS}

This coming week's APS Meeting has a large number of DGRAV-sponsored sessions. The plenary sessions related to DGRAV are \\

\noindent {\bf Plenary I: The Physics of Steven Weinberg} (Sat April 9, 8:30-10:18), particularly {\it Steve Weinberg's Universe} (8:30-9:06, {\it Speaker: John Preskill}) \\

\noindent {\bf Kavli Foundation Keynote Plenary Session: Update on Cosmological Parameters} (Mon April 11, 8:30-10:18). {\it Speakers: Wendy L Freedman, Jo Dunkley, Nikhil Padmanabhan}\\

Further, there are 14 invited and 27 parallel sessions this year.
A full list of all DGRAV-sponsored sessions can be found at this link: \href{https://april.aps.org/sessions?sponsors=DGRAV}{DGRAV sessions}.
I list the invited sessions here:\\

\noindent{\bf Next-Generation Gravitational Wave Observatories} (Sat April 9, 10:45-12:33, joint with DAP). {\it Speakers: Alessandra Corsi, Matthew J Evans, Emanuele Berti} \\

\noindent{\bf Holographic Developments in Quantum Black Hole Physics} (Sat April 9, 1:30-3:18, joint with DPF). {\it Speakers: David Lowe, Clifford V Johnson, Sebastian Fischetti}
\\

\noindent{\bf Beyond Einstein Gravity: Predicting Astrophysical and Cosmological Signatures} (Sat April 9, 3:45-5:33). {\it Speakers: Stephon Alexander, Helvi Witek, Jose Maria Ezquiaga}\\

\noindent{\bf Primordial Black Holes} (Sun Spril 10, 10:45-12:33, joint with DAP). {\it Speakers: Mairi Sakellariadou, Joseph Silk, Gabriele Franciolini}\\

\noindent{\bf 50 years of Black Hole Entropy} (Sun April 10, 1:30-3:18 pm). {\it Speakers: Rafael D Sorkin,Matthew Headrick, Andrea Puhm}\\

\noindent{\bf Soft Graviton and Gauge Amplitudes, Asymptotic Symmetries, and Bootstrap} (Sun April 10, 3:45-5:33, joint with DPF). {\it Speakers: Julio Parra Martinez, Daniel S Kapec, Monica I Pate}\\

\noindent{\bf Progress and Challenges in Analytic Gravity} (Mon April 11, 10:45-12:33). {\it Speakers: James A Isenberg, Daniel Kabat, Jared Speck}\\

\noindent {\bf Mergers of Neutron Stars: Nuclear Physics from Gravitational Waves} (Mon April 11, 10:45-12:33, joint with DNP). {\it Speakers: Sophia Han, Leo Tsukada, Tyler Gorda}\\

\noindent {\bf Developing Constraints on Quantum Gravity} (Mon April 11, 1:30-3:18, joint with DPF). {\it Speakers: Gary T Horowitz, Isabel Garcia Garcia, Migel Montero}\\

\noindent{\bf Astrophysical implications of GW190521} (Mon April 11, 1:30-3:18, joint with DAP). {\it Speakers: Michela Mapelli, Robert Farmer, Abbas Askar}\\

\noindent {\bf Isaacson Award Session} (Mon April 11, 3:45-5:33). {\it Speakers: Ling Sun, Aaron Zimmerman, Peter K Fritschel}\\

\noindent {\bf Binary Neutron Star Evolution and Post-Merger Physics} (Tues April 12, 10:45-12:33, joint with DAP). {\it Speakers: Sanjana Curtis, Daniel Siegel, Carolyn Raithel}\\

\noindent{\bf Challenges and Advances in Numerical Relativity} (Tues April 12, 1:30-3:18, joint with DCOMP). {\it Speakers: Manuela Campanelli, David Radice, Sherwood A Richers}\\

\noindent {\bf Black Holes and Gravitational Waves via Effective Field Theory Methods} (Tues April 12, 1:30-3:18, joint with DPF). {\it Speakers: Walter D Goldberger, Leonardo Sentore, Mikhail Solon}\\

\pagebreak
\section*{}
\vspace{-1cm}

\begin{center}
{\LARGE Cosmic Explorer, the Next Exploration Step}\\
\vspace{5mm}
Varun Srivastava\footnote{vasrivas@syr.edu} (Syracuse University),
        Kathryne J. Daniel\footnote{kjdaniel@brynmawr.edu} (Bryn Mawr College),
        Stefan Ballmer\footnote{sballmer@syr.edu} (Syracuse University)
\end{center}
\addtocontents{toc}{\protect\medskip}
\addcontentsline{toc}{section}{
{\bf Cosmic Explorer, the Next Exploration Step}, {\it by Varun Srivastava, Kathryne J. Daniel, and Stefan Ballmer}}

The current gravitational-wave detectors Advanced LIGO and Advanced Virgo have revolutionized our understanding of the Universe with only two years of integrated observation time.
Since first detection in 2015, the field of gravitational-wave astronomy has been transformed by 90 confirmed gravitational-wave detection events~\cite{abbott2021gwtc}. 
The observed sources are predominantly binary black hole mergers at a redshift of less than one~\cite{abbott2021population}. These events have served as a means to test general relativity and opened a new window for observational astrophysics and cosmology~\cite{abbott2021tests}.
Their progenitors span a wide range of masses and spins, probing the binary black hole population in today's universe and their formation channels in galactic fields and globular clusters. 
Gravitational waves from neutron star-black hole mergers and binary neutron star systems have also been detected~\cite{abbott2017gw170817}.
In particular, the counterpart of the binary neutron star merger GW170817 was observed with a wide range for astronomical instruments spanning the entire electromagnetic spectrum~\cite{GW170817_Joint}. These observations have advanced astrophysics with a better understanding of the neutron star equation-of-state, the dynamics of gamma-ray jets, kilonovae, and r-process nucleosynthesis~\cite{abbott2018gw170817, radice2018gw170817}.

Increased detector sensitivity enables us to extract the rich astrophysical information encoded in the gravitational-wave signals.
In the next couple of years current gravitational-wave detectors will approach their facility sensitivity limitations. Further improvement of existing facilities then becomes a game of diminishing returns.
Future progress in the field hinges on constructing a next-generation gravitational-wave facility.
The Cosmic Explorer Observatory will see about 300,000 binary neutron stars each year and will be able to detect stellar-mass binary black hole mergers to out to a redshift of 100.
These detectors will be the backbone of gravitational-wave astrophysics for the remainder of the 21st century.

\section*{The Cosmic Explorer Concept}
\begin{figure}
    \centering
    \includegraphics[width=\textwidth]{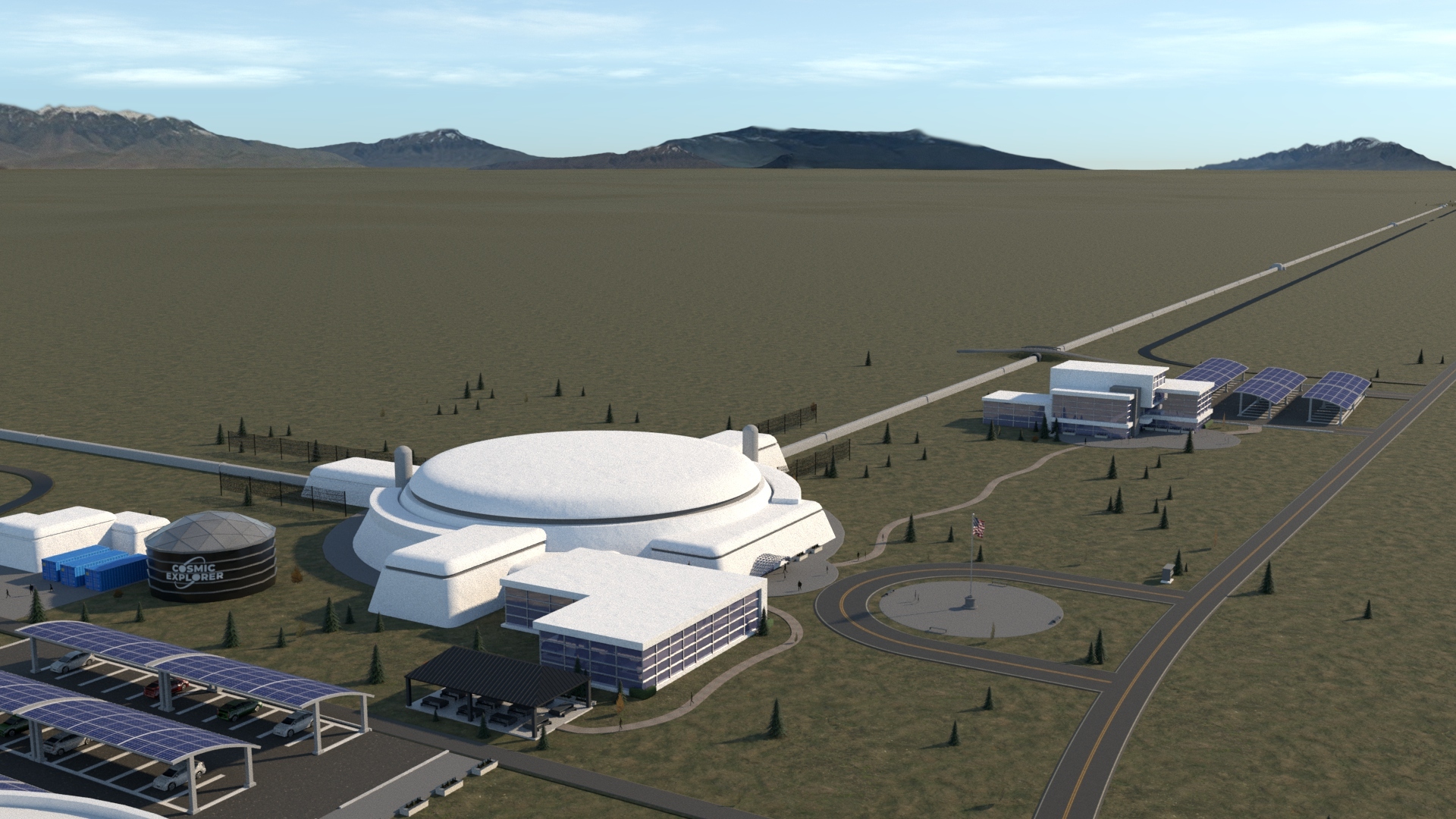}
    \caption{Artist's concept of the corner station at one of the envisioned Cosmic Explorer Observatory sites. (Credit: Eddie Anaya, Cal State Fullerton)}
    \label{fig:ce_art}
\end{figure}

The proposed Cosmic Explorer Observatory consists of two separate facilities in the United States, each with a single, L-shaped gravitational-wave detector~\cite{Evans:2021gyd}. It is intended to be operational by the mid-2030s. The concept was fleshed out in the NSF-funded Cosmic Explorer Horizon Study, which presents the key science goals of the Cosmic Explorer Observatory and compares the performance of different networks of detectors, optimizing the detector arm lengths within each network. 
The study revealed that at least two Cosmic Explorer detectors are necessary to meet all key science goals~\cite{Evans:2021gyd,borhanian2022listening}. The reference concept thus consists of two Cosmic Explorer detectors, each an L-shaped detector with equal arm lengths of 40~km and 20~km, respectively. Fig.~\ref{fig:ce_art} shows an artist's rendering of one of those envisioned facilities.

The Cosmic Explorer detector design is in many ways an evolution of the Advanced LIGO detectors.
Most fundamental noises scale down with the increase in length of the Fabry-Perot arm cavities. 
In the frequency range below 2~kHz a 40~km Cosmic Explorer is at least ten times more sensitive than Advanced LIGO, more at lower frequencies (Fig.~\ref{fig:strains}).
To further reduce quantum noise the Cosmic Explorer detectors will, compared to Advanced LIGO, operate at two times higher circulating power in the arm cavities, with eight-fold heavier test masses, and with improved frequency-dependent squeezing. Notably, the power density on the main optics will actually be lower, alleviating problems due to absorption and thermal distortion of the test masses.
To achieve the Cosmic Explorer design sensitivity above 2~kHz the losses in the signal extraction cavity need to be at least a factor of 10 lower than the Advanced LIGO detectors. 
This requires capabilities to sense and mitigate mode-match losses and reduce optical losses.

Seismic noise and thermal noise limit the low-frequency sensitivity of the detector.
The low-frequency sensitivity is crucial for the detection prospects of intermediate-mass black holes, and for providing an early warning of an impending merger.
The Cosmic Explorer reference design employs longer suspensions, better seismic isolation, and Newtonian noise suppression techniques to improve the low-frequency sensitivity~\cite{PhysRevD.103.122004}.
These improvements allow Cosmic Explorer to achieve more than a factor of 100 improvements below 20~Hz compared to the A+ design, see Fig.~\ref{fig:strains}.

\section*{The Science Goals of Cosmic Explorer}
\begin{figure}
    \centering
    \includegraphics[width=0.95\textwidth]{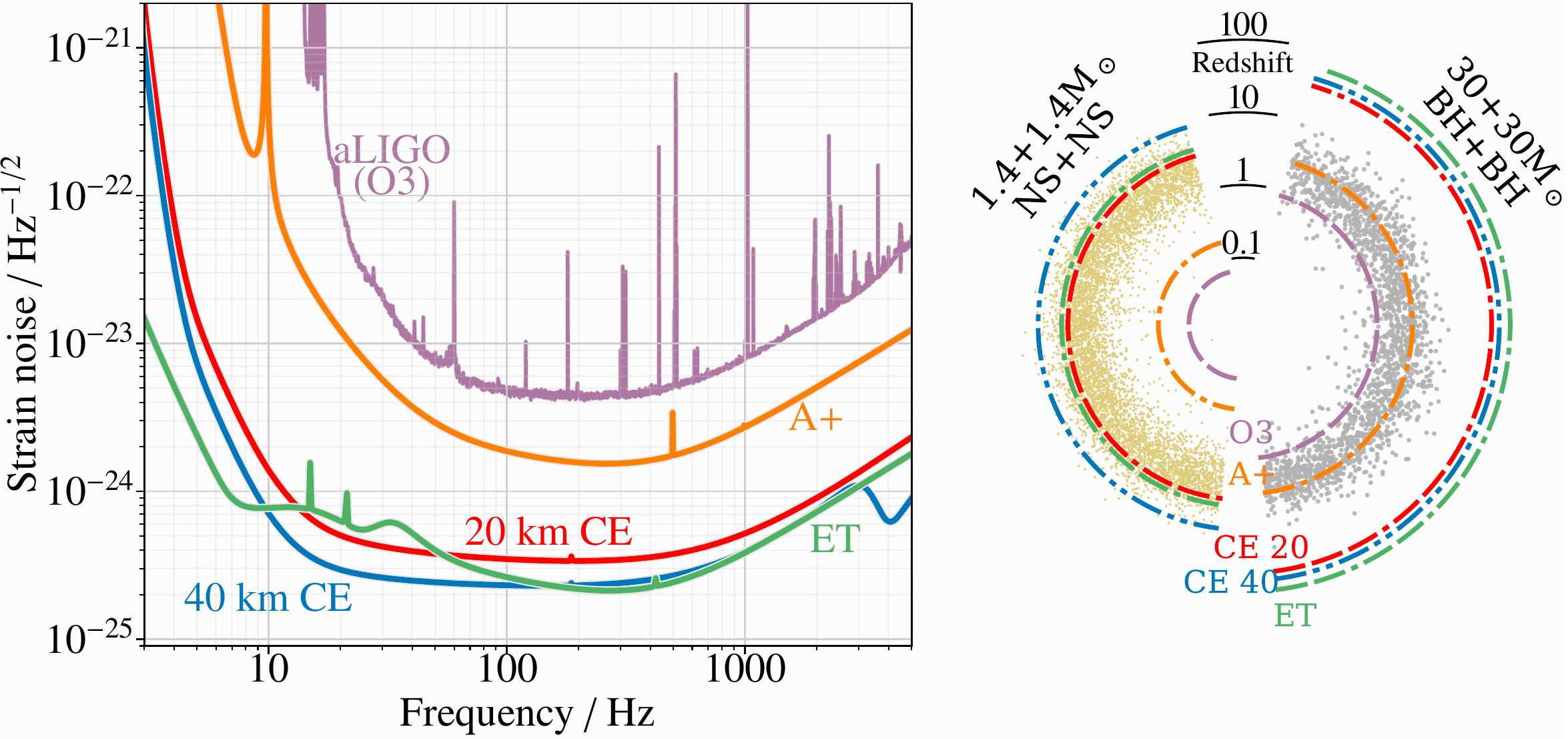}
    \caption{The left plot compares the sensitivity of the current 4~km long LIGO detectors with the next-generation gravitational-wave detectors: the 40~km- and the 20~km-Cosmic Explorer (CE) detector, and Einstein Telescope (ET).
    The right plot shows the ability of future gravitational-wave detectors to detect gravitational-wave sources from the entire population of binary black holes and neutron stars.
    }
    \label{fig:strains}
\end{figure}

The Cosmic Explorer Observatory will challenge the current understanding of astrophysics, nuclear physics, cosmology, and extreme gravity.
The improved sensitivity allows the Cosmic Explorer Observatory to detect stellar-mass black holes out to a redshift of 100, exploring regions of the cosmos unreachable with electromagnetic observatories.
The Cosmic Explorer Horizon Study recognizes three science goals that cannot be met with the current gravitational-wave detector network and will revolutionize the field of astronomy:

\noindent \textbf{Black Holes and Neutron Stars throughout Cosmic Time:} 
The direct consequence of the improved sensitivity of Cosmic Explorer is the ability to observe the very first black holes formed in the universe and almost the entire populations of binary neutron stars in the universe, see Fig.~\ref{fig:strains}.
The formation of supermassive black holes is an unresolved question in astronomy.
If the supermassive black holes form from a cascade of mergers of stellar-mass black holes, these mergers will be detected by Cosmic Explorer.
The Cosmic Explorer Observatory will answer this question complementary to the James Webb Telescope, which can observe the signatures of an direct collapse to a supermassive black hole.
Observation of the high-redshift tail of the neutron star population, accessible only with the Cosmic Explorer Observatory, will provide an insight into the formation mechanism of neutron star binaries in the early universe.

\noindent \textbf{Dynamics of Dense Matter:} 
Neutron stars are the densest known form of matter in the universe. The dynamics of matter at these extreme pressures and densities are yet unexplored. 
The neutron star tidal distortions in a binary are engraved on the gravitational-wave signals as they coalesce and merge, which probes the structure and composition of neutron stars and is critical to understanding the cold equation-of-state of neutron stars.
The remnant of a neutron star collision produces post-merger gravitational waves that probe the densest matter at finite temperature, a region of the quantum-chromodynamics phase space that is otherwise inaccessible in laboratory experiments.
The Cosmic Explorer Observatory will observe about 50 sources per year with a signal-to-noise greater than 200, which will tightly constrain the cold equation-of-state. For the strongest signals they will provide access to the post-merger signal related to the hot equation-of-state.
The multi-messenger follow-up of the gravitational-wave signals with the electromagnetic counterpart will facilitate the understanding of heavy element nucleosynthesis in the universe.
To facilitate follow-up observations the Cosmic Explorer Observatory can alert the electromagnetic telescopes up to five minutes before the merger of binary neutron stars at a redshift $z<0.2$ and localize these loud signals within 10 deg$^2$ in the sky. 

\noindent \textbf{Extreme Gravity and Fundamental Physics:}
Gravitational waves probe the laws of physics right up to the black hole event horizon. 
The loudest signals observed with Cosmic Explorer can have a signal-to-noise ratio of 3000 or higher.
The Cosmic Explorer Observatory will observe $\sim$10 binary black hole mergers each year with a signal-to-noise higher than 1000.
These signals will test the nature of gravity at extreme curvatures and allow the most stringent tests of General Relativity.
Any deviations from General Relativity will transform the understanding of extreme gravity and open an unfathomed window for fundamental physics.
Lastly, the large set of gravitational-wave signals observed with Cosmic Explorer opens a new window to probe the nature of dark matter and dark energy.

\section*{Contributing to Cosmic Explorer}
\begin{figure}
    \centering
    \includegraphics[width=\textwidth]{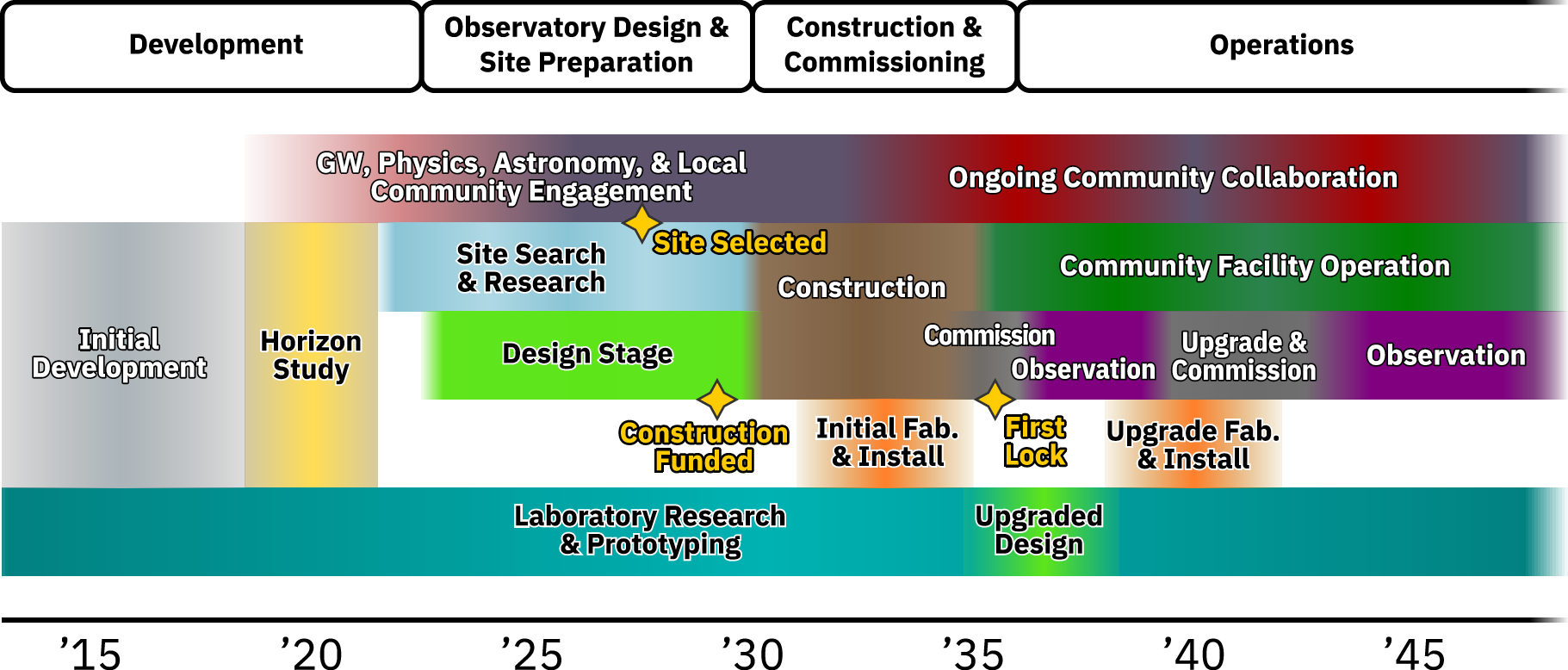}
    \caption{Proposed timeline of the Cosmic Explorer Observatory, envisioning the first observations to start in the second half of the 2030s.}
    \label{fig:ce_timeline}
\end{figure}
With the NSF horizon study finished, and with a strong endorsement from the Astro2020 Decadal Survey as well as the NSF DAWN~VI meeting, Cosmic Explorer is now transitioning from a concept to the project phase.
The immediate goal is to acquire funding for the conceptual design phase, as defined in the NSF's research infrastructure guide. 
While the organizational structure is expected to change during the project's design phase, the institutions behind the horizon study, MIT, Syracuse University, Penn State, CalState Fullerton, and Caltech, are currently taking the lead to maintain momentum and advance the project. 
They are pursuing the funding to initiate project-specific work, such as starting the site search and developing a project execution plan. 
At the same time, the window is wide-open for R\&D contributions from the entire gravitational-wave research community, and we expect the NSF to support this technology development effort through the traditional gravitational-wave research funding avenues.
In addition to the Cosmic Explorer project, the Cosmic Explorer Consortium has also been established. 
Its intend is to provide low-threshold access community to support the design, enabling technologies, astrophysics, and data science required for Cosmic Explorer to succeed.

\section*{The Road Ahead}

The envisioned timeline for Cosmic Explorer spans multiple decades and takes place in distinct stages:
development; observatory design and site preparation;
construction and commissioning; initial operations;
planned upgrades; operations at nominal sensitivity;
future observatory upgrades and operations, see Fig.~\ref{fig:ce_timeline}. Arguably the most time-critical task on this timeline is search, characterization and selection of  appropriate stretches of continuous land for the observatories. The facilities will have a significant impact on the landscape, environment and the local community. As such the local community needs to be involved in the development, and an assessment of the environmental impact must be included in the site selection process from the beginning.  The Cosmic Explorer team sees this as a novel opportunity to build mutually beneficial, long-term relationships with its host communities, particularly the Indigenous peoples of those lands, in the interest of caring for the land and its inhabitants, creating culturally supported pathways for the inclusion of Indigenous people in STEM, and ensuring the success of the Cosmic Explorer experiment.  

With observations planned to start in the mid-2030's, the Cosmic Explorer observatory will drive a significant growth in gravitational-wave astronomy, astrophysics and cosmology. In collaboration with the European Einstein Telescope it will take gravitational-wave observations back in cosmic time to the remnants of the first stars.

\bibliographystyle{unsrt}
\bibliography{MoG55.bbl}

\end{document}